\documentclass{CUP-JNL-DAP}%

\usepackage{graphicx}
\usepackage{multicol,multirow}
\usepackage{amsmath,amssymb,amsfonts}
\usepackage{mathrsfs}
\usepackage{array}
\usepackage{amsthm}
\usepackage{titlesec}
\usepackage{csquotes}
\usepackage{apacite}
\usepackage{rotating}
\usepackage{appendix}
\usepackage[authoryear]{natbib}
\usepackage{ifpdf}
\usepackage[T1]{fontenc}%
\usepackage{times}%
\usepackage{newtxmath}%
\usepackage{textcomp}%
\usepackage{xcolor}
\usepackage{hyperref}
\articletype{RESEARCH ARTICLE}
\jname{Data \& Policy}
\jyear{2023}
\titleformat{\paragraph}[hang]
{\normalfont\normalsize\bfseries}{\theparagraph}{1em}{}
\setquotestyle{british}
\SetBlockThreshold{1}

\begin{document}

\begin{Frontmatter}

\title[Pathogen-related data sharing during a pandemic]{COVID-19: An exploration of consecutive systemic barriers to pathogen-related data sharing during a pandemic}

\author*[1]{Yo Yehudi}\email{yochannah.yehudi@postgrad.manchester.ac.uk}\orcid{0000-0003-2705-1724}
\author[1]{Lukas Hughes-Noehrer}\orcid{0000-0002-9167-0397}
\author[1]{Carole Goble}
\author[1]{Caroline Jay}

\authormark{Yo Yehudi \etal}

\address*[1]{\orgdiv{Department of Computer Science}, \orgname{University of Manchester}, \orgaddress{\street{Oxford Road}, \postcode{M13 9PL}, \country{United Kingdom}}}

%TC:ignore
\abstract{
%250 words max - Abstracts should be 250 words. It must be able to stand alone and so cannot contain citations to the paper's references, equations, etc. An abstract must consist of a single paragraph and be concise. Because of online formatting, abstracts must appear as plain as possible.
In 2020, the COVID-19 pandemic resulted in a rapid response from governments and researchers worldwide - but information sharing mechanisms were sometimes sub-optimal. At the time of writing in 2023, while spread tracking has largely ceased, cumulative confirmed deaths approach 7 million, and cumulative excess death estimates range from 18 to 32 million. Modern vaccines reduce severity, but immunity wanes fast, and the virus mutates so swiftly that some individuals have contracted multiple cases. All observed infection cases - even ones where the acute phase is “mild” - result in invisible cardiovascular damage, even in previously healthy people, and repeated infections increase risk of the long-term consequences known as “Long COVID”. 

Despite this staggering toll, those who work with pandemic-relevant data have faced significant systemic barriers to accessing, sharing or re-using this data. In this qualitative study we interview data professionals working with COVID-19-relevant data types, analyse our interview data for content and themes, and report the barriers, successes, and opportunities for improvement. 

We identify four sequential barriers a researcher may encounter: Knowing data exists, being able to access that data, data quality, and ability to share data onwards. A fifth barrier, human throughput capacity, is present throughout all four stages. Examples of these barriers range from challenges faced by single individuals, to non-existent records of historic mingling / social distance laws, and up to systemic geopolitical data suppression. } 

\begin{policy}
% 120/120 words - Provide a 120 word statement here that summarises the significance of the work for policymakers, written at a level understandable for a broad audience...
Preventing the spread of a global pandemic requires effective, swift access to high-quality and up-to-date data. Despite this urgency,  data are often stored behind access-controlled systems that prevent effective re-use, and human time constraints for access requests result in bottlenecks, especially in high-demand pandemic-instigated scenarios. Even when data can be accessed, poor quality often makes it difficult or impossible for researchers to effectively use this data, and may necessitate lengthy workarounds or "best guesses" to create policy-informing computational models. To reduce costly death tolls in the future, we must implement effective computational data-sharing pipelines and permissive re-use policies. This paper provides new information that can be used to develop data-sharing policies that will support effective responses in future pandemics.
\end{policy}
%TC:endignore
\end{Frontmatter}

\section{Introduction}
Barriers to effective use of data to inform policy are not new, but became noteworthy during a global pandemic that has already infected hundreds of millions (\cite{owidcoronavirus}). Building on data-sharing challenges we encountered as scientists practicing in our own fields, we designed a qualitative remote interview-based study to investigate these barriers. 

After exploring literature (section 3), we present methods (section 4), and then share a comprehensive list of the consecutive systemic barriers our participants encountered when working to address the effects of the pandemic (section 5). This list is presented to inform policy and create effective sharing mechanisms before another such deadly event occurs. 

Whilst this research was intended specifically to look at non-private data, such as hospital bed capacity, or viral genomes, many participants had experience with private data types as well - and it quickly became clear throughout the study that most of the barriers we identified were common to both private and non-private data types. Much of the evidence we present are quotes and use-cases of the barriers our participants encountered, and given the sensitivity of topics such as data privacy (section 5.1.2) and governmental data suppression (section 6.1), all participants and quotes are anonymous. 

The barriers reported fall into five main categories, four of which are sequential: Knowing data exists, accessing that data, being able to understand and use the data, and being able to share your data and analyses in an online repository so others can re-use them. Throughout this, a fifth barrier is interwoven: Human throughput capacity, which may be stretched even before acute stress periods such as a pandemic.

This study focused specifically on recruiting people who wish to share barriers that they had encountered. Whilst it is not the primary focus of the study, we also discuss common themes and "wishlist" items that our participants shared, when describing what a good - or even "dream" data source might look like (section 5.3).

In the discussion (section 6), we highlight three key areas of note: the governmental tension between following science and controlling pandemic narrative; equity concerns as those with the least resilience are also the worst affected; and the need for structured computationally-readable temporal metadata for legal rulings such as COVID social distancing requirements. 

Finally, we conclude the paper with a short summary and call to action (section 7). 

\section{Research Question}
% see interview guide.txt for interview question set
In times of a pandemic or epidemic when rapid response is required, what are attitudes towards pathogen-related data sharing? In particular, what barriers do researchers encounter, and what do they do, if anything, to get around these barriers? 

\section{Related Literature}

Literature related to this study's goals spans a number of topics. We conducted a targeted literature review, specifically covering domains that were addressed by study participants or that emerged as recurring barriers during the interviews. Articles were retrieved by a combination of recommended reading from participants and domain experts as well as Google Scholar searches for the keywords "data sharing policy", "privacy vs transparency", and "open data", in combination with modifiers such as "COVID" and "healthcare" when a narrower scope was required.

In particular, we have reviewed literature related to: governmental transparency and data sharing; tensions between the right to personal privacy in medical records, vs. sharing data for public good; COVID-19 data sharing policies; secondary data use and re-use; industry and research institute data sharing policies, and data characteristics that make data useful (including the FAIR principles). 

\subsection{Personal medical privacy vs. collective greater good} 
Throughout this paper we do not dispute that individuals have a right to confidentiality in health. The tension between sharing private medical data for public good and the right to personal privacy is well documented in literature, even outside pandemic and epidemic-related concerns.  Acknowledgement of this right goes back to at least 320 BCE, if not earlier, with Hippocrates' vow not to divulge what he heard in the course of his profession \cite{jones_hippocrates}.  

\cite{Leslie_McSwain} observe that pandemic-related health data sharing presents a specific urgency: \enquote{In an emergency setting, we need to drastically reduce the barriers to frictionless [Health Information Exchange]: one law of the land for COVID-19–related data.} They further propose to ease common anonymisation measures in the face of pandemic need, such as reducing the threshold of "anonymous" geographical location blocks from 20,000 people to a smaller range of 50 to 100 individuals. 

\cite{henderson_patient_2021} notes that the right to individual privacy in epidemics could reasonably be abrogated to provide better population-level epidemic response, much like right to free movement is restricted during lockdowns, but eventually concludes that the benefits are not likely to outweigh the risks, given that these datasets are not likely to be representative and bias free, and risk being exploited by commercial entities. 

Biomedical data needs to draw from varied sources in order to provide meaningful and timely responses. \cite{leonelli_2023}, highlights that attitudes and expectations around evidence-based medicine resulted changed as a result of pandemic urgency. In previous decades, randomised control trials (RCTs) were more likely to have been used by researchers over all other data sources. Data were often biased to easily digestible or computation-friendly data sources, rather than "complex disaggregated data sources". Leonelli highlights the need for varied and multidisciplinary data sources to create a whole and realistic picture of epidemiology. These sources should include social factors which are likely to influence outcomes and behaviours, linked data (where possible) and data sources from medical professionals and social workers, rather than only from researchers. 

\subsection{Transparency and data sharing in local and national governments}

Effective government inter-organisational data sharing provides expertise, revenue, policy, and efficiency benefits (\cite{gil-garcia_government_2016}, \cite{ramon_gil-garcia_collaborative_2007}). Given the inherent risk and tension around personal privacy vs transparency, it is not surprising that government data sharing policies vary not only from country to country, and local authority to local authority, and even between individual governmental agencies in the same country. Governmental data are often siloed, however clear direction from high-level leadership can ease the silo culture and make inter-agency consensus and collaboration easier (\cite{graham_navigating_2016}). 

Governmental administrative data are often limited by practical constraints. \cite{allard_state_2018} list some of these: internal lack of capacity to analyse it; data that is of poor quality that needs significant preparation before it can be analysed, and a lack of common data formats and identifiers. Whilst IT infrastructure and data-management systems are necessary to solve these issues, they are far from the only element needed. Factors that can improve these socio-technical challenges include early intervention when or before data sharing problems arise, favourable political environment and internal policy and active data sharing culture-setting, especially from organisational leadership (\cite{dawes_need_2009}). \cite{gil-garcia_government_2016} finds that both having a project manager dedicated to inter-organisational data sharing and having sufficient financial resources significantly correlated with governmental inter-organisation data sharing success.  \cite{allard_state_2018} further observes that data associated with mandatory compliance reports tends to receive more attention and resource than other data types, and are often of higher quality.

\cite{allard_state_2018} and \cite{graham_navigating_2016} further note that researchers are well-positioned to lead with use of data systems, and as such it would be advantageous for governments to collaborate with researchers in policy-forming partnerships. These partnerships are shaped not only by government agency perception of risks, but also by the type of data access structures in use, suggesting that careful design for sharing early on is more likely to be effective. 

Researchers may be disincentivised from working with governmental agencies due to the time taken to navigate data-access bureaucracy, especially if there is no standardised process to access the data and/or if data access agreements forbid re-sharing data.  

\subsection{Transparency, data sharing, and secondary data re-use in academic settings}
A sub-set of academic researchers have been pushing for openly shared research data for some time - for example the Bermuda principles for genomic data sharing were established in 1996 (\cite{maxson_jones_bermuda_2018}).

Generally, whilst opinions on research and academic data sharing are favourable, in practice many researchers find that the effort associated with data sharing (i.e. preparing data to be shared and/or metadata that contextualises the raw data) and risks (ethical risks, misinterpretation, risk of scoop, lack of incentive) often outweigh an individual’s reasons to share data (\cite{Yimei_Zhu_Open_access_in_uk}, \cite{datasharing_sociology}). It is thus perhaps unsurprising that in a survey of 1,800 UK-based academics by \cite{Yimei_Zhu_Open_access_in_uk}, 86\% of respondents rated data sharing as "very important" or "fairly important", but only one fifth of the respondents had personal experience of sharing their data. 

 It takes time and effort to prepare research data for sharing, both making sure that the data are tidy and documented, as well as making sure that the data are sufficiently contextualised that they are not dangerously misinterpreted (\cite{datasharing_rcts}, \cite{Yimei_Zhu_Open_access_in_uk}, \cite{empirical_datasharing_plos}). Whilst many journals require data availability statements when publishing scholarly results, data availability is rarely enforced by the journals. In replication and data sharing studies for journals that have data sharing expectations, \cite{empirical_datasharing_plos} were able to access only one dataset out of the ten they tried to access, while \cite{datasharing_rcts} successfully gained access to data for 19 out of 37 studies they wished to replicate. \cite{datasharing_sociology} note that only a small number of research journals state how data sharing policy adherence is checked, and further note that there is a gap between recommendations from policy makers and the realities of data sharing in social sciences. 
 
 In addition, data sharing practices vary by research domain. \cite{Yimei_Zhu_Open_access_in_uk} observed that natural sciences and engineering are more likely to share data than medical and life sciences, whilst \cite{datasharing_rcts} notes that while biomedical sciences generally do not share data, its sub-domain of genetics does have a data sharing culture. \cite{datasharing_sociology} notes that half as many social science articles had data available compared with political science journals, observing that this may be down to specific ethical concerns in each domain - sociology is more likely to have qualitative interview data that cannot be effectively anonymised, compared with standardised quantitative political survey data. 

Infrastructure to support data sharing is a final important element. \cite{Yimei_Zhu_Open_access_in_uk} notes that humanities data sharing is not only blocked by privacy concerns, but also by a lack of infrastructure on which to share the data, and \cite{Kim_Zhang_data_repos} found that when people had data repositories available to them, they were more likely to actually act on their good intent to share data.

\subsection{Data standards, integration, quality and usability}

One observation in the previous section on academic data sharing is that preparing data for re-use takes time and effort. It is therefore unsurprising that there are many calls for data standardisation and reviews on the (often less than ideal) current data standards compliance in many domains.  

\cite{fairchild_epidemiological_2018} observe that epidemiological data are challenged both by a lack of standardised \textit{interfaces} to data, and by a variety of data \textit{formats}, leading to fragmentation. Different access mechanisms span from computational Application Programming Interface (API) access, to manual human extraction from Portable Document Format (PDF) files. Formats may range from general and simple data formats such as comma separated value (CSV) files, which are easily readable but can vary massively in terms of content, to more complex dedicated domain-specific time series formats such as EpiJSON. They also observe that just because a standard exists does not mean it is necessarily embraced by the field, and whether or not a given data format will be widely adopted is often an ongoing question. 

Even when standards exist, they may not meet needs of their users. \cite{gardner_need_2021} notes that in 2020, systems for COVID-19 data reporting \enquote{were not empowered or equipped to fully meet the public's expectation for timely open data at an actionable level of spatial resolution} - that is, whilst systems for sharing this data existed, they didn't provide the depth of information needed to be useful. 

In genomics and bioinformatics, \cite{thorogood_international_2021} observe that data may be standardised in a single data source (e.g. a particular research organisation or lab), but vary across different data sources of similar types. This results in adherence to standards that nevertheless makes federated (cross-data-source) queries a difficult technical challenge. \cite{thorogood_international_2021} proposes that an independent organisation such as the Global Alliance for Genomic Health (GA4GH) can act as a convener for its member organisations to help create interoperable data standards. 

Building on collective experience of research data management, \cite{wilkinson_fair_2016} launched the FAIR data principles, which assert that data should be Findable, Accessible, Interoperable, and Re-usable (FAIR) if it is to effectively support research discovery and innovation. To be FAIR, data should be indexed on a publicly accessible data portal, use a common data standard and/or vocabulary, provide machine-readable data access and detailed descriptive metadata that is available even if the original data are not available, and provide clear data provenance and licence information about how the data can be re-used. 

Similar principles have been described for other data-relevant domains, such as the FAIR software principles (\cite{chue_hong_fair_2021}), which provide similar guidelines for the computer code that produces, analyses, and visualises datasets. The European Commission with ELIXIR has established a pan-European open COVID Data Portal (\cite{elixirCovidPortal}) for public data deposition and access, and hosts an open letter with over 750 signatories, calling for open and FAIR data sharing (\cite{openDataLetter}). Additional COVID-19 FAIR data recommendations can be found in \cite{FAIR_data_for_a_coordinated_COVID-19_response}.

\section{Methods and data used}
To gain nuanced understanding of the barriers to data access, sharing, and re-use, we conducted a qualitative study, interviewing professionals who work with COVID-19 related data. This study was approved by the University of Manchester Department of Computer Science ethics panel, original approval review reference 2020-9917-15859 and amendment approval reference 2021-9917-19823. 

A detailed interview guide and study protocol is available on protocols.io for re-use (\cite{covid_study_protocol}).

\subsection{Recruitment}
Interviews were carried out in two rounds, and all participants who applied to participate were accepted, resulting in fifteen completed interviews. 

Round-one (July-September 2020 and March 2021) recruitment was targeted at bulk-emailed mailing groups and slack channels, with additional recruitment via social media (Twitter). No potential participants were approached directly individually, and everyone who volunteered to participate was accepted into the study. Areas approached were primarily COVID-19 hack events, open data communities, and bioinformatics / biology communities. 

Due to the limited global spread of the round-one sample, we initiated a second round of recruitment. Round-two (July-September 2022) recruitment was more directly targeted towards specific individuals who had domain experience in non-private data, and/or were based in continents and countries that were previously not represented in our sample. Round-two participants were approached via instant messaging and email. 

Due to the low level of response in round-one, in round-two, we did not focus on any specific domain beyond experience of working with COVID-19 data in some way, resulting in the sample domains described in results section 5.1.

\subsection{Data gathering and analysis}

\subsubsection{Data gathering: interviews}
Prior to interviews, participants were sent an information sheet via email explaining the purpose of the interviews and how their data would be handled. 

Interviews were semi-structured, conducted remotely via an institutional Zoom\cite{zoom_2022} video conferencing account. We guided participants through onscreen consent forms  before asking the following questions: 

\begin{enumerate}
\item Tell me a little about the data sources you've been working with so far – have any been restrictive or hard to access?
\begin{enumerate}
\item What did you do when you encountered the access restriction?
\item Did you ever consider going to another source, or encouraging data creators and curators to share their information somewhere less restrictive?
\item Did others you work with share your views and opinions? Can you provide any examples?
\end{enumerate}
\item What about good experiences – are there any data sources that are exemplary? 
\begin{enumerate}
\item What did they do right?
\item Was there anything they could do better?
\end{enumerate}
\item If you could design your own “dream” COVID-19 data source, would there be anything
else you’d like to see?
\item Are there any ethical issues you’re aware of, or have considered, with regards to this
data sharing?
\item Is there anyone else working in this domain who you think might be interested in
participating in this study?
\end{enumerate}

\subsubsection{Data management and analysis}
Post-interview, recordings of the interviews as well as transcriptions were downloaded from Zoom, speech-to-text errors were corrected, and then the text transcriptions were analysed for themes. Transcripts were coded in NVivo 12 \cite{qsr_international_pty_ltd_nvivo_2018}, resulting in 304 codes in total. Of these, we filtered a list of 101 codes that were present three or more times across the fifteen transcript files. A second coder reviewed the filtered list to confirm the presence or absence of codes in each transcript file. 

Of the 1,515 data points (101 codes * 15 files), both coders agreed for 1,514 of the codes (greater than 99.93\% agreement). After discussion between the coders, agreement was reached on the final code, giving 100\% agreement for all codes present three or more times across the transcripts. 

All codes that appeared three or more times relating to barriers are presented in section 5, with illustrative quotes where possible. All quotes used in the text are shared as a supplemental data file. 

\textbf{Subject matter was potentially sensitive:} participants may have spoken about disapproval of data sharing practices in their institutions or countries, and/or may have witnessed, conducted, or experienced inappropriate or illegal data sharing activities. All interview videos were deleted after transcription. To preserve anonymity, corrected interview transcripts were assigned a pseudonym identifier and stored securely on University of Manchester infrastructure for at least five years in line with University of Manchester guidelines. Transcripts are not deposited in an open repository. 

\subsubsection{Preserving anonymity in reporting}
When writing, quotes are in some cases lightly paraphrased to replace identifying details, e.g. a phrase such as "Transport for London" might be replaced with the phrase "[municipal transit authority]". All participant quotes that are re-used in publications were first shared with the participants, who were given the chance to amend or redact their statements to avoid inadvertent data breaches. This approach has been applied in various cases to place names, institutions and governmental / healthcare agencies and committees, and data source names. 

To mitigate against the potential re-identification of the individuals whose data we share, we have created multiple separate tables for datapoints such as location, occupation, and domain expertise. For example, whilst we might share a table of occupations of participants, and separately share a list of countries where they come from, we never link occupation and physical location data of participants in the same table, as a combination of specific datapoints may be sufficient to re-identify an individual.

\subsection{Reflexivity and positionality statement} 
This research was carried out by researchers based in the United Kingdom, and led by YY. Whilst YY has lived in various high-income settings around the world, including New Zealand, the United States, and Israel, they also have high levels of experience working with scientific data and running open-focused equity-driven global scientific communities. Most recently, this includes working with \textit{OLS} (Formerly Open Life Science), an open science mentoring and training organisation, as well as software engineering and data manipulation experience at \textit{InterMine}, an open source biological data warehouse. 

In particular, we recognise that the study was designed to focus on \textit{open} data, by a researcher who advocates for and prefers to work with open data whenever possible and ethically prurient. Importantly, we do not assert that this should be taken as a comprehensive or balanced review of data access mechanisms - instead, this study specifically investigates \textit{barriers}, and how professionals may handle them. 

In the sample section, we also discuss limitations of the sample and recruitment method - as noted above, our communities are open-focused and equity-focused, which is likely reflected in the sample we were able to reach for interview. 

\section{Results}
\subsection{Sample}
Fifteen participants were interviewed in total, through two recruitment rounds. Six participants signed up during round one, recruited via bulk-mailing lists and social media. All round-one participants were researchers based in Europe or North America, and came from a range of domains spanning healthcare and computational, biological and social research.

Round two recruitment was more directly targeted towards individuals, and resulted in an additional nine participants, for a total of fifteen participants overall in the study. Round two brought participants from Africa, Asia, Australasia, and South America, with two thirds of the final sample based in High Income Countries (as defined by the World Bank) and one third based in Low or Middle Income Countries. 

Tables 1 to 3 show profiles of the participants. These results are intentionally dis-aggregated to present a profile of participants without linking the data in a way that might facilitate de-anonymisation.

\begin{table}[ht!]
  \begin{center}
    \caption{Participant location}
    \label{tab:locations}
    \begin{tabular}{r|l} % <-- Alignments: 1st column left, 2nd right, with vertical lines in between
      \textbf{Region or country income status} & \textbf{no. participants} \\
      \hline
        High Income Country (HIC)&10\\
        Lower or Middle-Income Country (LMIC) &5 \\
        \hline
        Europe&7\\
        North America&2\\
        South America&2\\
        Asia&2\\
        Africa&1\\
        Australasia&1\\
    \end{tabular}
  \end{center}
\end{table}

\subsubsection{Domain Expertise}

Participants came from a broad range of overlapping professions and expertise domains, shown in \textit{Table 2} below.  The number of participants adds up to greater than fifteen due to some participants falling into more than one category.  

\begin{table}[ht!]
  \begin{center}
    \caption{Participant domain expertise}
    \label{tab:professions}
    \begin{tabular}{r|l} % <-- Alignments: 1st column left, 2nd right, with vertical lines in between
      \textbf{Domain} & \textbf{no. participants} \\
      \hline
      Computational biology / bioinformatics (non-clinical)  & 4\\
      Clinical patient informatics & 4\\
      Civic/governmental data sharing & 3 \\
      Mobility data & 3\\
      Epidemiological modelling & 1\\
      Journalism & 1\\
      Software engineering (non-biological) & 1\\
      Mathematics & 1\\      
      Social media research & 1\\      
      Gender and race inequality & 1\\      
    \end{tabular}
  \end{center}
\end{table}

\subsubsection{Data sensitivity: Private, non-private, commercial, and political data}

Pandemic data falls on a spectrum from the individually identifiable, such as patient medical records, data which may be identifiable depending on the level of detail, i.e. mobility data and hospital admissions, and to data where there are no individual privacy concerns, such as viral genome sequences.  

Throughout the paper, we will work with the following definitions: 

\textbf{Private data} refers to records that, if shared inappropriately, could compromise an individual's right to privacy, such as medical records, detailed geolocation data, or personal human genomes. 

\textbf{Semi-private data} refers to data records that do not have direct identifying information such as names or addresses, but that nevertheless may be re-identifiable depending on the level of detail, such as phone mobility data, census-based commute information, or hospital admissions.

\textbf{Non-private data} refers to records that do not compromise individual privacy as defined above. This might include hospital bed capacity information, vaccine availability, or viral genome sequences.

We also define two categories of data that fall across the privacy spectrum: Commercial data and Political data.

\textbf{Commercial data} are data that are withheld from the public due to industrial competition and intellectual property reasons. This category spans both private and non-private data above - for example, a for-profit science lab might decline to share a viral genome it has sequenced, considering it proprietary sensitive information (which would be defined in this paper as commercial and non-private), or a health insurance company might consider information it could learn from its health records to be proprietary (which in this paper we would define as both commercial and private data).

\textbf{Political data} are data that may or may not be private, but that a government agency of some kind wishes to hide from the public view. In the case of this study, all data that fell in this category tended to be data that made the pandemic look severe in some way, e.g. hospitals over capacity, high community / school COVID-19 case loads, or deaths. 

\paragraph{Data sensitivity: profile of study participants}

Despite this study initially aiming to recruit participants who worked in non-private subject domains, many participants who signed up worked with private and semi-private data sources as well. These participants were not excluded as many of the access, quality, and re-sharing barriers were still relevant to the research question. As Table 3. shows, most participants worked with more than one of these categories of data. 

We neither targeted nor disqualified participants based on their experience with commercial or political data. 

\begin{table}[h!]
  \begin{center}
    \caption{Participant data sensitivity expertise}
    \label{tab:data_sensitivity}
    \begin{tabular}{r|l}  % <-- Alignments: 1st column left, 2nd right, with vertical lines in between
      \textbf{Data sensitivity category} & \textbf{no. participants} \\
      \hline
      Private data & 10\\
      Semi-Private data & 10\\
      Non-Private data & 10 \\
      Commercial data & 9\\ 
      Political data & 4\\ 
    \end{tabular}
  \end{center}
\end{table}

\subsubsection{Data types: profile of data types discussed by participants}

\begin{figure}[ht!]
    \centering
    \includegraphics[width=1\linewidth]{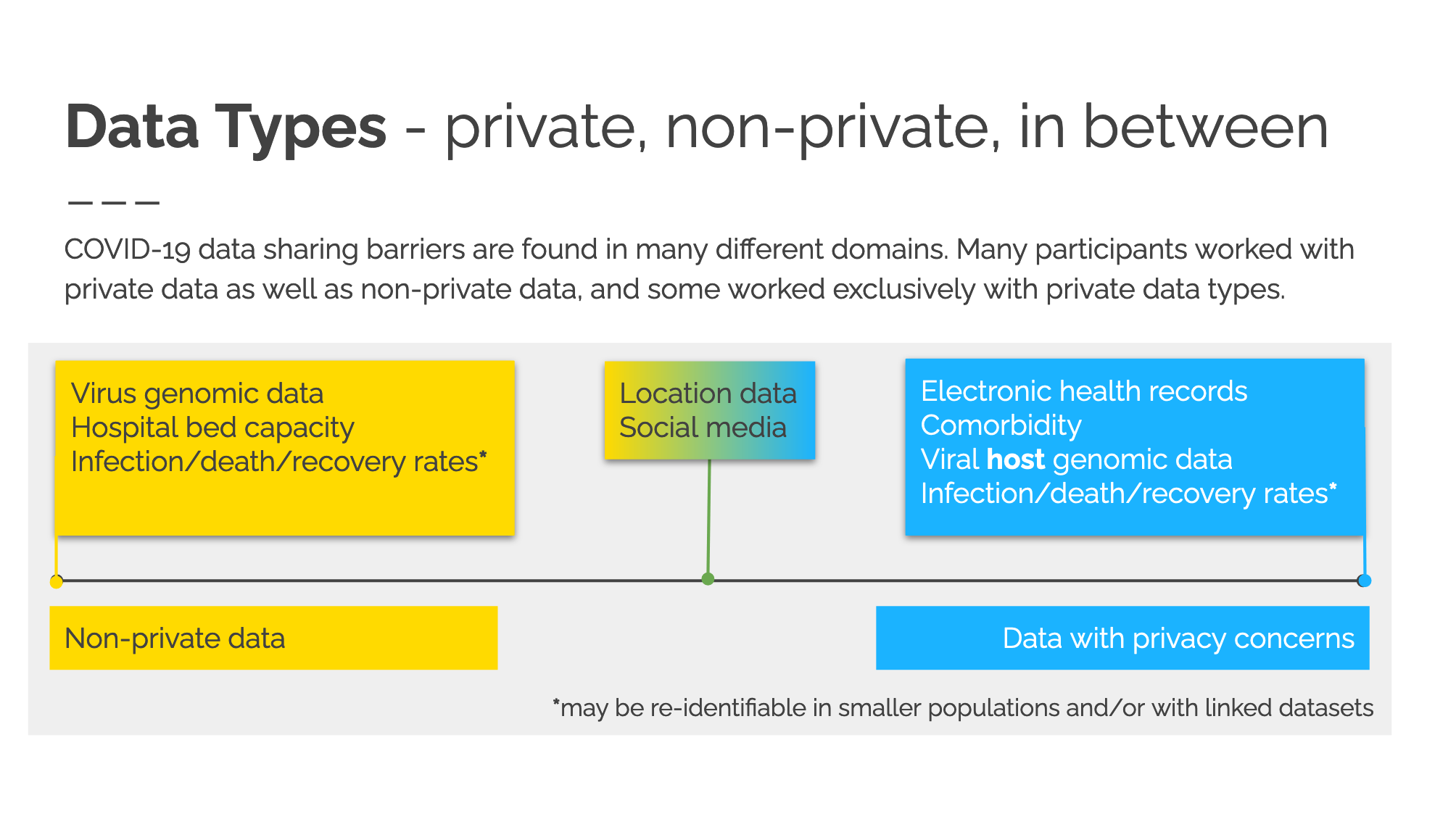}
    \caption{Data types in the study covered a spectrum ranging from highly personal and private, to data that would not threaten personal privacy if shared}
    \label{fig:fig-datatypes}
\end{figure}

Figure \ref{fig:fig-datatypes} illustrates the spectrum of private and non-private data discussed in this study. The study was initially designed in response to barriers the authors had encountered around non-private data: viral genome and COVID-19 spread information. 

In the course of the interviews, the scope of data expanded to cover additional data types: 

\textbf{Non-private, but potentially commercial and/or political:}
\begin{itemize}
\item Genomic virus sequences.
\item Vaccine supply.
\item Vaccines administered.
\item COVID-19 cases (tests and positivity rate) - in the general community and in schools.
\item COVID-19 hospitalisations.
\item COVID-19 and excess deaths.
\item Hospital bed availability - intensive and regular beds.
\item Hospital deaths (with or without COVID-19 positivity, as excess deaths are still pandemic-relevant).
\item COVID-19 related lockdown and interpersonal mingling rules.
\item Geographical regions and population density.
\item Synthetic data designed to resemble real private data.
\item Metadata and data structures for private data, without the private data itself. 
\end{itemize}

Viral sequences are potentially commercial if sequenced by organisations that wish to exploit the sequence for commercial gain, and can be political as well. More than one participant noted that there were viral sequences from Wuhan, where the pandemic originated, that later disappeared from the servers they were on. 

Vaccine, hospital, and COVID spread data in particular are data types that were political and may have been obfuscated, redacted, or simply not tracked by government or health officials. This is further explored in the discussion section of this paper.

\textbf{Semi-private}
\begin{itemize}
\item Mobile phone mobility data - actual geolocation, exercise routes, or planned navigations using maps. Commercial. 
\item Any data from the non-private category, when it concerns individual people and the data available is in very low numbers and/or combined with additional datapoints that could de-anonymise individuals. 
\item Race, age, gender, sexuality, ethnicity, disability, socioeconomic status.
\item Census commute data (place of work and place of dwelling). 
\item Social media.
\end{itemize}

\textbf{Private}
\begin{itemize}
\item Patient medical records (anonymised, pseudonymised, or not anonymised).
\item Viral host / patient genome records.
\item Any of the semi-private data above when very detailed and re-identifiable, even if it is pseudonymised.
\item Lockdown related domestic abuse, femicides. 
\end{itemize}

\subsubsection{Sampling challenges and limitations}

This study is not intended to be generalisable and comprehensive, but instead it is exploratory, designed to shed light on areas for potential future action, exploration, and research. 

All participants who reached out to participate in the study were accepted, regardless of background, which resulted in a population that was skewed towards higher-income European and North American countries. Ideally sampling would have been more evenly balanced across continents, cultures, and expertise domains. Because of the small sample size, we did not segregate the analysis by any of these features. 

We speculate that there may be several possible reasons for the sample distribution:

\begin{enumerate}
\item \textbf{Recruitment methods}: Initially ethical consent for this study was granted on the condition that no individuals were approached directly to participate, as this direct approach may have seemed too coercive. Whilst we emailed international mailing lists, Slack groups, and had thousands of engagements with recruitment tweets, uptake was still very poor. Round two had a much higher uptake rate, after an amendment to the ethics plan was approved which allowed researchers to directly approach potential participants - but was then limited to targeted individuals within the researchers' networks. 
\item \textbf{Time constraints}: Several potential participants expressed interest in participating but did not ultimately sign up or attend the interview slot, and one participant was originally contacted during round one but eventually only signed up to participate in round two, citing extreme busyness. Given that people who have expertise in pandemic-related data sharing were generally involved in pandemic response, it seems likely that many potential participants self-selected out of participation due to high workload.
\item \textbf{Personal and reputational vulnerability}: Given the sensitive nature of the subject matter, people may have been reluctant to participate for fear of re-identification and subsequent responses. Even people who \textit{did} consent to participate were cautious about the information they shared. Multiple participants redacted some of their statements when given the option, and/or indicated regret for having been as candid as they were in the interview. One participant even went so far as to change their Zoom display name and turn off their camera before they gave consent to record the interview. 
\end{enumerate}

\subsection{Barriers}

\subsubsection{Barrier 1. Knowing the data exists, and being able to find it}
In order to work with or request data, researchers and data analysts must know the data exists. Multiple interview participants talked about their experiences gaining access to non-private data that was neither deposited on a public data repository, nor made use of, nor publicised. Indeed, one participant describes accidentally stumbling onto restricted-access but highly relevant bed capacity data whilst browsing an institute shared data archive: 

\blockquote{It was only on an off-chance in perusing through these documents on a Sunday evening. I was having a chat with my colleague, and we're both going \enquote{what else exists in this data dump that we could do something useful with?} when we stumbled across [this datasource].}

Even if data are made accessible to the public in some way, e.g. by being deposited in a data repository that is not access controlled, discoverability can still be a challenge, as data are usually distributed across a broad range of repositories and may not be designed to facilitate data discovery. Three separate participants reported that they had manually curated lists of data sources on a word processor document because there was nowhere else that effectively collated relevant data sources. Participants share their experiences finding data: 

On finding data within a datasource with minimal metadata:

\blockquote{They're literally just sorted by date. PDF only. No underlying data, and unless you know what the [document] was about, it is impossible. If you say \enquote{what does it say about travel restrictions?}, unless you know when they met to discuss that, good luck finding it without literally going through 300 PDFs.}

On finding out whether or not viral sequences exist anywhere accessible:

\blockquote{There are lots of viruses and sequences that fall under the radar [...] when you don't have the data in a central repository. So that's been kind of the linchpin. A lot of people are working on, probably the same viruses, and it's not that we don't want to cite them, it's just when you're doing a high throughput analysis it's hard to track down every single case and really do your due diligence that someone didn't say \enquote{oh this library actually has this sequence, and we named it.}}

And on discovering data and data attributes in publicly accessible data: 

\blockquote{Unless you know things are available, you can't ask for them, so it's not obvious that you can get data by sex and age from the dashboard. [...] I've got this document I have collected over the last 18 months, and it's just links. It's got everything. New cases types of test, variants of concern, age profiles, countries, contact tracing data, international comparisons, positivity rates, public health. It's a 10 page document now, and I still add new things to it. What's frustrating is that you kind of have to know where these things are, otherwise you can't really access it. Although it's all public data, it is not accessible, not really, right?}

Even once data has been found, it may not be easy to find again if the naming conventions are not logical and there is little metadata: 

\blockquote {It's going to sound obvious, but more metadata would be great. Every time I need to transfer between [two tables] I spend longer than I should searching the website. It's surprising how easy it is to lose one of these tables, and you can't find it again because they have funny names.}

\subsubsection{Barrier 2. Access: Barriers to obtaining data}
The next hurdle researchers face is gaining access to data sources they have identified as relevant or useful. While many participants in this study focused on access to non-private and non-clinical data, we still found that researchers faced systematic access barriers. 

\paragraph{Cost}
Cost can range prohibitively from thousands to tens of thousands or more for government agency-hosted data, and for mobile-phone generated mobility data. Participants reported that some providers shared data without additional cost - for example, Facebook (https://dataforgood.facebook.com/dfg/covid-19) and Google (https://health.google.com/covid-19/open-data/) shared their mobility data publicly without charging for it. For-pay data prices ranged from sufficient to cover infrastructure and maintenance costs, up to commercial rates in the tens of thousands of pounds, euros, or dollars. One participant remarked that even "at cost" prices can still be significant when dealing with extremely large datasets. 

\blockquote{We've allocated for one year, something like £30,000 pounds to pay to get data extracts, which just means that when we're an unfunded volunteer effort - we can't. Whose grant does that come out of? […] I think it's just a business model. That's part of [government agency], that's how they pay their bills, to maintain their servers.}

Mobility data providers in particular varied based on which organisation was providing data. One participant reports:  

\blockquote{I think Three in Ireland shared their data initially for free, and then, after a couple of months, they charged cost for that content, to collect and publish datasets and share this data with the central government. [...] it's not expensive. Low four figures or less, for complete access. Whereas O2 and Telefonica expected, I think, upward of £10,000 pounds for the same data and they offered no discounts [...] they basically said \enquote{look on our website: we haven't changed anything during the pandemic, you've always been able to pay for this data with ten grand.}}

\paragraph{Culture} 
Two-thirds of participants (ten out of fifteen) referenced culture as a barrier around data sharing and access. Sometimes access control was treated as a default without questioning if the data actually \textit{needs} privacy-preservation measures.  Participants reported this both from an industry angle, where corporations were striving to keep intellectual property secret, as well as from academia, where non-private data were access controlled by default, by placing it on gate-kept institutional servers.

On individual researchers wishing to share data when their institute does not encourage it: 

\blockquote{Even people who I would say, \textit{want} to do this [data sharing] - they're just not going to do it - institutional momentum, right? \enquote{that's not how things are done and therefore that's not how we're going to do it}, and I feel like that's what's carried over from the 80s and 90s in this field, and they just never stopped to really reflect on this, \enquote{Is this the way that we really \textit{want} to do things moving forward?}}

And when a supervisor discourages data sharing: 

\blockquote{If the person who is overall responsible for organizing the whole thing doesn't seem in general to be concerned with open research or open data, or anything like that [...] at the point where there's a supervisor who's saying \enquote{this is how it's done} or not saying how it's done... I guess you don't want to step out of line.}

\paragraph{Institutional access barriers} 
In some cases, the type of institution or mode of employment may affect access to relevant data. One researcher reported that whilst public universities were allowed to access national COVID-19 related data in their country, any non-government-sponsored universities were not permitted to access the same data, even though they too might be using the data for the exact same purposes as public institutions. 

Data access agreements may not always be agreeable for a research institute, either: 

\blockquote{I would really like to use Facebook data for that, but I can't get my university to sign the agreement because it gives Facebook the right to use their branding in advertising.}

\paragraph{Volunteer Labour} 

Taking advantage of volunteer labour can also present institution-related access issues: Volunteers are not usually considered members of an organisation even if they are working on an institute-backed project. One participant reported needing to arrange honorary positions at a university for their volunteers, whilst another participant was asked to analyse data in a volunteer capacity, but then denied access to the same data (including access to the metadata model, which would not reveal any private data) by the same institute that had asked for help. Volunteer-related access barriers may also be technical, e.g. if the volunteer does not have access to an institution's computer network, they may not be able to connect to the institution's servers to access the data.

\paragraph{Friends in high places} 
One unevenly distributed data access route that participants mentioned was having the right connections. Multiple participants noted that knowing someone who was already involved in the project, or knowing someone extremely prominent, was often one of the most effective ways to gain access to data. 

One participant, reflecting on prominent people asking for data: 

\blockquote{If the mechanism for accessing data is \enquote{email someone and wait for them to respond}, then if you don't ideally have both professor and OBE involved in your name, you know you're going to struggle.}

Another participant, reflecting on gaining access to data due to incidental association with another researcher: 

\blockquote{We got lucky - we were grandfathered in because we sit within an adjacent group to the person who sat on the original […] committee. There was no formal application process.}

\subsubsection{Barrier 3: Utility: Hard-to-use data, even once available}
If a researcher successfully gains access to privileged data, or uses open data, they face a new challenge. Access alone does not guarantee data is well-documented or easy to understand. As one participant asserts: 

\blockquote{Essentially, you had a treasure trove of information that was not at all mapped to each other, that a lot could have been done with, which was being heavily access managed and not at all curated.}

\paragraph{Data trustworthiness} 

The biggest barrier by far in this category - reported by fourteen of the fifteen participants - related to trustworthiness of the data they were using. Often this was down to \textbf{poor quality control} - information that conflicted with itself, such as three PCR tests for a single patient in a single day: one negative, the second positive, and the third negative again. Other examples of poor quality control reported were:

\begin{itemize}
\item Obviously corrupt data, such as English sentences embedded in data that purported to be amino acid or nucleotide character sequences. This type of sequence usually consists of a limited sub-set of the alphabet, and usually appears to be random to the human eye - for example, the first few characters of the nucleotide sequence for BRCA1, a breast cancer gene, are "gctgagacttcctggacgggggacaggctgt". \cite{intermine_authors_brca1_2022}, \cite{smith_intermine_2012}.
\item Lack of outlier or input error correction, such as hospital bed capacity dropping from 50 to 0 overnight, then returning to 50 the next day. 
\item Variations between input from different people in the same setting. 
\item Incorrect reporting dates: for time series data such as vaccinations, infections, hospitalisations, and deaths, there may be reporting lags resulting in higher or lower numbers than the reality on a given date.
\end{itemize}

\paragraph{Data may have explained or unexplained gaps}

Gaps in data may arise due to causes such as intentional institutional or governmental redaction, human recording error, lack of time to record variables. Lack of foresight or planning can result in a failure to capture certain data aspects that might be useful including inpatient health-related variables, or socio-economic status, gender, ethnicity, and disability variables. 

\paragraph{Data provenance and methods may be missing}
Being able to understand where data came from (provenance) and how it was gathered (methods) is an essential part of being able to rely on data and use it for trustworthy analyses and policy / behavioural advice. 

Lack of provenance can result in data being discarded or ignored, even if it would otherwise have been useful. One participant illustrates a scenario where they know a datapoint, but are unable to provide the provenance of this knowledge:

\blockquote{Having access to highly protected things, like healthcare data, for half of my work is bizarre, because for example, we need to know what proportion of people who enter the hospital over the age of seventy dies? What is that rate? I know it, because I generated it for one of my models, but I can't use it for the other model.}

Multiple participants discussed issues with "derived" data, where a number or dataset is provided, but the methods to calculate these numbers are not clearly laid out, and can be difficult to trust or re-use. Often, this kind of barrier can result in laborious reverse engineering, and re-calculations based on guesswork. 

\blockquote{[There are] many pages of supplementary where we have documented everything we've done, because what's amazing is that I don't think the way [a healthcare organisation] calculates those numbers is necessarily right. But given that we're reverse engineering, it wasn't something we could call out, because we don't know the definition, but we're like \enquote{this doesn't make sense.}}

\paragraph{Spreadsheets, comma separated value files, and Microsoft Excel}
Multiple participants discuss issues with spreadsheets, mentioning CSV file issues, and sometimes expressly mentioned Microsoft's Excel data file format.

\blockquote{It is in an Excel sheet, which is kind of fine, except [...] the table appears partway down the sheet, with some explanatory text, and the order of the sheets changes, sometimes. [...] It just has a series of different tables copied one after the other in one CSV file, and you just have to know which line the table you want starts on, so if they ever reorder them it completely breaks.}

Whilst spreadsheet use facilitates data analysis and creation for non-data professionals, inexpert use can cause challenges. One participant reported that whilst their institution had a database-backed data repository, clinicians were extracting spreadsheet-based reports and editing/\textbf{annotating them manually in the spreadsheets, with no effective way to bulk-import the data back into the institutional database}. When data analysts were called in to create advanced data analysis pipelines, they then were forced to either sanitise and re-integrate the spreadsheet-based data sources, or abandon the meaningful additions created by clinicians. Longer term, this institute was able to identify and dis-incentivise fragmented records, but this requires both training and technical functionality to allow sufficient free-text or structured annotation. 

Another participant reported on spreadsheets used for \textbf{time-series data}, reporting on multiple healthcare institutions at a national level. Each day's data was represented in a single sheet. In order to create time-series based analyses, multiple sheets needed to be amalgamated across a long period of time, a highly technical challenge that might have been avoided by a more suitable data format. Institutional reporting lags and missing days further complicated this scenario. 

Spreadsheets facilitate \textbf{"hidden" columns, rows, and sheets of data}. This may be used to appear tidy, or to prevent inappropriate data editing, but when a human attempts to run data analysis, these hidden subsets of data may obscure important details, including meaningful data provenance.

One participant commented on the UK's COVID-19 case misreporting due to an .xls spreadsheet file failure: 
\blockquote{The other thing that this pandemic revealed [...] is that our data infrastructure just is not up for this kind of task, at least in the health sector. [...] I recall a case from from the UK, where essentially the government was reporting [...] most of the cases as an Excel sheet and at some time you know an Excel sheet was not going to be enough. I mean maybe a different kind of database could have done it. And you wonder like \enquote{well, hang on, I mean this is the way we're counting these in one of the most advanced countries on earth? So, well, what's up?}}

\paragraph{Temporal changes and temporal context}
Many participants pointed out that data often needs to change over time as columns are added, more efficient formats are created, or new analyses are devised. However, when a data format changes over time, this context may not be recorded, nor may the date of the change be noted. 

\paragraph{Geographical region definitions} 
Different geographical data sources may name the same regions differently - e.g. "Cambridgeshire" and "Cambs" are equivalent to human readers but stymie a computer script. Regions also change over time as boundaries and districts are re-defined, making it hard to reliably define regions without significant temporal context. Discussing difficulties converting from one geographical coordinate set to another, one participant stated 
\blockquote{Guides on how to do that from the government come with a big warning sheet saying you should under no circumstances use this for any software that informs policy, which is not very helpful... because we have to.}

\paragraph{Temporal contexts for mobility and behavioural changes}
Geography is not the only area that is enriched by temporal context. When interpreting infection spread rates, it is important to understand what legal movement and behaviour controls were in place at the time, and \textit{when} significant changes are made, such as lifting of lockdowns or social distancing. One participant reported that whilst it is usually straightforward to find what a given legal requirement is \textit{right now}, finding out what legal protection measures were in place for historical events becomes much harder. 

\paragraph{Inability to integrate or link datasets}
Sometimes different datasets have the potential to meaningfully answer a question present in the other dataset, but nevertheless cannot be combined. This barrier may be technical - i.e. there is no common unique identifier with which to harmonise records across data sources - or it may be a sociolegal barrier: some licence terms prevent data remixing. One participant illustrates how this can be a challenge.

\blockquote{What I don't want to see happen is to create like a multi tiered system or a system where data is not centralized, because that's the nightmare of bioinformaticians, where you have to go to some private database for these sequences, and then you go to the Genbank for these sequences and then cross referencing them, where are they duplicated...}

\paragraph{Data may not be designed for computational analyses}

Participants reported accessing data as Excel "data dumps" - potentially hundreds of sheets that must be laboriously manually downloaded by a human, or as PDFs, which are not easily read by a computational script. Other participants reported having to copy/paste data from web pages on a daily basis and re-typing data embedded in graphics. When data are not designed to allow programmatic access (e.g. via an API), fetching data and slicing it into the correct sub-sets of data (for example, fetching only hospitalised patients over a certain age), becomes difficult or impossible. 

\paragraph{Technical barriers} 
Many of the use barriers discussed thus far are a result of data structure design issues, training needs, and human error. Technical infrastructure, especially infrastructure that is suddenly used in unanticipated ways, may present additional barriers - it may not be designed to be accessed externally, e.g. because of use of a data safe haven, or it may not be designed to cope with a heavy load of users. One participant reported that they regularly experienced incomplete / interrupted downloads, potential rate limits, and had little choice but to script repeated attempts until a full data download was completed. 

\blockquote{[The] server just terminates at some point, and your curl command just hangs for hours or days. I mean, we had no workarounds in place [...], and resubmit data hundreds of jobs a hundred times until they succeed.}

\blockquote{If you have a pandemic, and if you have data sets that […] thousands of researchers on a daily basis want to use, or need to use, to create dashboards - our current infrastructure was struggling.}

\paragraph{"Data cleaning" is time consuming} 

Given the rife data quality and usability issues, it is perhaps unsurprising that participants reported laborious processes to prepare the data for analysis. One participant stated: 

\blockquote{The data came in a very basic format and we found a lot of issues. We spent a lot of time, in fact almost about a year, just to clean up the data.}

\subsubsection{Barrier 4: Further distribution: Barriers to re-sharing data and analyses}

Once a researcher overcomes the barriers in finding data, accessing it, and preparing it for use, they may face difficulty disseminating the analyses they have created - perhaps due to data use agreements that prohibit re-sharing data, or permit re-sharing of analyses and derived numbers, or simply require pre-approval before analyses are shared. 

\paragraph{"Open" data, but no sharing onwards}

Restrictive data sources may disallow any redistribution of their data, which can result in analyses being impossible to reproduce or researchers simply abandoning the datasource entirely in favour of less restrictive data sources. Participants weigh in on data sources with strict or prohibitive re-use policies: 

Two participants weigh in on GISAID, a viral genome sequence database: 
\blockquote{The GISAID database which is containing the SARS-CoV-2 sequences is blocked off, and it's kind of not worth the hassle to use that data, even though it might be informative, because if we use it we then can't share our data, which is a derivative.}

\blockquote{One of the things we care about, is not sharing \textit{just} the data itself, but also all the code and scripts and stuff that we use to analyze the data and produce the final output. But for the parts where we're building trees to compare the final genetics of our isolates versus the world's, we can't actually bundle that data together. The GISAID data that is used as the context is not distributable, and so that's where there's kind of this gap between our ideal of being able to to provide everything bundled together.}

\paragraph{Re-sharing only approved items}
Other data sources may require that they pre-approve all published items that use their data in some way:

On Strava Metro, a city mobility datasource, which has a policy requiring all re-use to be approved by Strava:

\blockquote {I spoke to one person from local government who just went \enquote{we're just not going to use it then} - there's just no value to it if you're going to provide that amount of restriction.}

Two participants talk about getting permission to share analyses of data that were deemed politically sensitive during the peak of a pandemic wave: 

\blockquote{A big battle that we had with [the data creator] was that [the data user] had made an agreement that, before anything was published, it would be run past them and so as we put our  preprint on [a preprint server], [the data user] let [the data creator] know, and there began a little six week ordeal where we would just go back and forth being like \enquote{can you please sign this off} and they'd be like... they didn't even respond, and then after six weeks, the only reason we got signed off was because they released a portion of the [dataset] into the public domain as they normally would.}

\blockquote{I think they've managed to publish a few papers, but they're always kind of outdated because it takes so long for [governmental body] to say yes. And they weren't allowed to preprint them, so I know that that has been quite frustrating and would have been actually really important data to help shape how people talk about COVID-19 and adherence.}

\paragraph{Data integration results in redistribution challenges}
A recurring theme researchers reported was mixed-permissions datasets are hard to re-share. Data users are forced to choose whether to ignore restrictive datasets entirely, and use only open-permissioned datasets, or to integrate richer restricted datasets but have difficulty distributing their work. Mixed-permission datasets can either be redistributed with gaps, sharing only the open elements and making their research non-reproducible, or alternatively, data distributors may invest in time- and resource-heavy multi-tier access systems. 

\paragraph{Sharing may be blocked due to data quality concerns}
Whilst ensuring quality is unarguably important, over-strict rules can prevent even good quality data for being shared onwards. One participant reports that despite their data being suitable for BLAST, a popular computational biology analysis tool, their re-shared data was deemed "third-party" (i.e. less trustworthy), and wasn't stored in a way that permitted BLAST analysis: 

\blockquote{A bit of a snag with [a genomic sequence database]. [The database] was like \enquote{Oh, we'll take the sequences, but then they're going to be classified as third party assemblies}, which means that they're not BLASTable, they're not really essentially useful in any real way. And so now I need to try to convince [the database] that \enquote{no, these are viral genomes} - they might not be complete in every case, but these are very distinct viruses and there is no representation in [the database].}

Another participant describes the \textbf{trade-off between making data easy to share and enforcing quality}: 
\blockquote{Both go hand in hand, and the challenge is - and I don't have a good answer for that - how can we keep good metadata, but still make it easy to upload? [...]  We are now at a stage where we have an Excel table, so people can use their tools that they know and put data in, and then we upload it with an Excel table. But still, it still can be more complicated than [another database that the participant expressed quality concerns with]. So yes, there's a trade off. But I guess, if we do our best, and if we make this as simple as possible, people will hopefully do the extra effort of the metadata.}

\subsubsection{Barrier 5: A barrier woven throughout: human throughput}

Human throughput creates bottlenecks throughout the entire process of working with data: it takes time and effort to find existing data sources, humans are gatekeepers for access control, data are often recorded manually by busy staff, many data pipelines and analyses require intervention to run, and human dissemination is required to share results. 

\paragraph{No time to record data or respond to access requests}
People working in these areas may have been resource-stretched even before COVID-19 created acute demand periods. A participant comments on pandemic response staff having trouble finding the time to accurately record patient information: 

\blockquote{All this data is manually updated by the residents who ran helter-skelter across the various COVID-19 wards, and whenever they get time, they do it, so we can not expect much. But fairly by and large, it's okay.}

Multiple participants commented that people often had too much going on to handle data access requests: 

\blockquote{I would have loved the other stuff, but everything they had of value was being either guarded, or people were just too focused on doing the work to be able to facilitate others getting access.}

\blockquote{Lots of people were working flat out and it's just a time and resources barrier for being able to… lots of people in different bodies were working at capacity, and therefore just did not have time to do extra stuff.}

\paragraph{No time to try new approaches or analyses}
In other cases, a team may have access to data but not have enough time to try new things or to analyse the data they have, or quality of data produced may be poorer than in non-crisis times.

One participant elaborates on an un-used data set: 

\blockquote{[They were] given a real big data set and it sat untouched by the people on the [analysis] team. It was really exciting kind of access and they could have probably done quite a lot with it, but they were just working on the things that they always worked on in a way that they always worked on. They didn't have time to go \enquote{Okay, how do we actually integrate this into the data we already collect in the dashboards, that we already produce?}}

Another participant comments on trying to encourage their colleagues to produce better-quality data: 

\blockquote{They don't exactly understand why they have to, and felt lost and even misunderstood - like a reaction of \enquote{come on, I have millions of things to do before, and you came and are telling me that I'm not doing my work well}, you know? No, because I'm gathering data in excel spreadsheets...}

\subsubsection{Barriers: A summary}

Table 4 concisely presents key points from each of the five barrier types. 
\begin{table}[ht!]
    \centering
    \begin{tabular}{ | l | l| }
       \hline
       \textbf{Barrier} & \textbf{Key points} \\
       \hline
         % BARRIER 1 %
         \multirow{4}{10em}{\textit{Barrier 1: Knowing data exists and being able to find it.}} 
          & Data exists only on \textbf{private hard disk} or \textbf{private repository}.\\
          & \textbf{Un-indexed} data - no way to search for an individual data point.\\
          & Disparate \textbf{data sources not collated}, so easy to miss existing sources.\\
          & Technically open data, but \textbf{access mechanisms not public or easy to find}. \\
         \hline
         % BARRIER 2 %
          \multirow{5}{10em}{\textit{Barrier 2: Access: Barriers to obtaining data}} 
         & \textbf{Cost} to access or purchase data is prohibitive. \\
         & Organisational \textbf{culture} discourages sharing.\\
         & \textbf{Institutional requirements} make it hard or impossible to sign agreements. \\
         & Volunteer labour - \textbf{no official contract} with the institute that owns the data. \\
         & Lack of "\textbf{friends in high places}" to expedite data access requests.\\
         \hline
         % BARRIER 3 %
         \multirow{9}{10em}{\textit{Barrier 3: Utility: Hard-to-use data, even once available}} 
         & Poor data quality control / \textbf{low data trustworthiness}.\\
         & Incomplete data, with \textbf{gaps}: \\
         & -- Redaction, including government-level \textbf{data suppression}.\\
         & -- Missed recording or lack of time to update records.\\
         & \textbf{Missing provenance} and missing collection \textbf{methods}. \\
         & \textbf{Inappropriate file formats}, e.g. Excel, images, PDF.\\
         & \textbf{Missing \textbf{temporal} context} for data that changes over time:\\
         & -- Geographical boundaries redrawn over time.\\
         & -- Legal changes over time. \\
         & Inability to \textbf{link datasets}. \\
         & \textbf{Computational infrastructure under-powered} / under-resourced. \\
           
         \hline
         
         % BARRIER 4 %
          \multirow{4}{10em}{\textit{Barrier 4: Further distribution: Barriers to re-sharing data and analyses}} 
         & Data may be openly accessible, but \textbf{sharing \textit{onwards} is disallowed}.\\
         & \textbf{Re-sharing} potentially allowed, but \textbf{only for approved items}. \\
         & Data \textbf{integration} results in \textbf{redistribution} challenges.\\
         & Sharing may be blocked due to \textbf{data quality concerns}. \\
         
         \hline
         % BARRIER 5 %
         \multirow{3}{10em}{\textit{Barrier 5: A barrier woven throughout: human throughput}} 
         & No time to \textbf{record data}. \\
         & No time to respond to \textbf{access requests}.\\
         & No time to learn, \textbf{try new approaches} or new analyses.\\
         \hline
    \end{tabular}
    \caption{Key points from each of the five barriers}
    \label{tab:barriers-summary}
\end{table}

\subsection{Good experiences and "dream" data sources}
Whilst presenting the many consecutive barriers to effective data sharing in a pandemic can appear bleak, there were also beacons of light. Participants described both specific datasets and aspects of existing data sources that worked well, as well as suggesting "dream" or "wishlist" items that may not have all existed, or at least not all in a single place, but that participants considered desirable aspects of a high-quality data source. Here, we will describe the attributes that participants valued, and when there are no anonymity concerns, we will provide the example dataset names that illustrate a given point. 

Both the wishlist and the examples of existing good data sources covered areas that addressed barrier 1 (knowing data exists), barrier 2 (accessing data), and barrier 3 (hard-to-use data). No wishlist items or good examples specifically addressed barrier 4 (re-sharing data) or barrier 5 (human throughput). 

\subsubsection{Addressing Barriers 1 \& 2: Actively sharing or updating data} 
Barriers 1 and 2 are around knowing data exists, and then being able to access that data, so it is perhaps unsurprising that pro-actively sharing data was a behaviour that made people particularly appreciate dataset creators and originators. Participants applauded organisations - research organisations, government sources, and commercial organisations - which pre-emptively shared their data without needing to be asked.

One participant discusses un-prompted sharing of viral genomic sequences: 

\blockquote{The Craig Venter Institute had tons of Sanger sequencing on coronaviruses, and in January as soon as the pandemic started, they released everything into the SRA, and they didn't make a big deal about it or anything - they were like \enquote{here are all these coronaviruses that we've had sitting in our database.}}

Another participant pointed out that \textbf{reducing data access application overhead} when possible can make a difference. Here, they reference a Facebook-provided dataset: 

\blockquote{They have shared data that usually you would need a very lengthy agreement to get. You do have to sign the agreement, but it's like \enquote{tick this box on one form.}}

Other examples that were cited included a dataset which, without prompting or being asked, \textbf{was updated to comply with newer data standards}, and the \textbf{public release of Google, Apple, and Citymapper records from mobile phone mapping} / navigation records. 

In a similar vein, one participant reported the value of \textbf{knowing \textit{when} a dataset had been updated}:

\blockquote{When the numbers are moving fast, they send us updates like two or three times per day. This is really nice, and we can actually have an idea of what's going on, like on a daily basis.}

\subsubsection{Addressing Barrier 3: Making data easy to use} 
The third barrier we presented was around the difficulties of using data, once (or if) a participant has successfully found and accessed the data they needed. Easy-to-use datasets as described by study participants were machine readable, integrated multiple data sources and data types seamlessly, were trustworthy - they didn't need pre-processing, verification, or "cleaning" - and they had standardised tooling available to wrangle the data with.

\paragraph{Integrated and/or linked data}
Participants wished frequently for data that was integrated - creating a scenario where disparate datasets from different sources can be treated as a single large dataset, and queried as one. 

One participant cited two projects, PIONEER and DECOVID, that attempt to integrate data in this way, and explain how useful this could be in UK-based medical scenarios. More info on these datasets is available via \cite{pioneer}

\blockquote{Trying to bring together two complete electronic health record data sets in OMOP [an electronic health record data standard] is a non trivial task, I don't know of anyone that had actually done it prior. [...] It's attempting to [...] create an end to end the solution, where from the moment you interact with acute services, to when you end up in critical care, to your discharge [...] every person that interacts through the NHS 111 [National Health Service health advice hotline] but then went to A\&E [Accident and Emergency Department] and ended up in critical care, then got discharged - and being able to match it to ONS [Office of National Statistics] outcomes, so that you know, actually, who died, not just the partial observation that we get in medical records - that would be the dream, but I think that's everyone's dream.}

\paragraph{Computational data access and "big data"}
Large datasets can be hard to download - both challenging in terms of time and in terms of data storage. \textbf{Mirroring large datasets online} in multiple places may allow a data user to do their processing work entirely "in the cloud" - bringing the computational analysis to the data, rather than bringing the data to a local machine: 

\blockquote{A good example for me in terms of datasets - the Thousand Genomes data set is a really nice one because the data is completely open.  We now have high-quality versions of that dataset that we can host on cloud [...] one of the nice aspects is you can find mirrors on various platforms, so that regardless of what infrastructure you're using, it's usually easy to access it and not have to do a lot of work to get a copy, because as these data sets get bigger and bigger you really start running into limitations of how reasonable, how feasible, it is to actually get a copy for yourself.}

\textbf{Programmatic access to data} - that is, making it easy for a computer script to access the data without any human intervention needed - was cited as a key attribute of Genbank and RefSeq: 

\blockquote{Genbank, RefSeq have been the anchors for this project, and that's just purely based on their availability policy. I can grab everything, I can search it, I have programmatic access to get the right data. Which is fundamental, because when you're dealing with 40,000 entries you can't manually go through this much data. I mean, we still have to go through the data, but you can't slice it efficiently without programmatic access, right?}

Programmatic access could be further enhanced by creating highly predictable computational querying schemas, using an accession number (unique computational identifier) and clear URL structure: 

\blockquote{All the data ends up being migrated into a data warehouse where you can predict the URL of every single object, as long as you know the accession [...] that not only allowed us to build a website around database and data, but that also lets anyone access the data that they want just by typing in the URL.}

\paragraph{Wishlist: Open and standardised tooling}
Making data easier to use also involves making sure there is useful tooling available to manipulate the data \textit{with} - to allow people to clean, slice, filter, and visualise the data they are working with.

One participant wished for open source tooling - that is, tooling where the computer code for the tool is freely available to access, and the tool itself is generally available at no monetary cost to its user: 

\blockquote{Maybe such Open Source solutions could be made available, so others could implement at no cost, and this can become more commonplace, so analysis could be done and more can be extracted from that data to bring benefit back to the patients from whom the data came.}

Fake or "synthetic" health records that match the exact data fields of real health records allow health record analysis to be carried out computationally. A coder can design the analysis and test it on the fake data, without ever having to access privileged private health records. Someone who \textit{does} have access to the records could then run that analysis on the real document set. One participant wishes for standardised tooling to create datasets like this: 

\blockquote{To generate synthetic data, to generate fake data, there are also starting to be tools in place, but there are no standards yet. This is another blocker, because if you need to produce fake data, you have to figure out how. You don't have a protocol or procedure.}

Another participant wished for tooling (and high-quality data) to allow pandemic predictions - not for modelling human movement and behaviour, but for modelling what effects different viral genome mutations have. 

\paragraph{Wishlist: Trustworthy data sources}

A wish for trustworthy data sources was another aspect that mirrored the frustrations of barrier 3. Desired aspects of trustworthy data include "clean" data that doesn't need preparation before being used, data in standardised formats where possible, the ability to annotate existing data sources with third-party annotations, and knowing what biases may exist in your data.

\textbf{Data standards:} participants reference the desire for having clear naming conventions for entities, and creating schemas for data presentation .

\textbf{Third-party data annotation:} One way to facilitate data trustworthiness is by allowing data users to share their knowledge and experience using a given data source - such as by creating machine-readable third-party annotation options for data sets: 

\blockquote{I think annotations of datasets are missing [...] third party annotations need to increase here. [...] \enquote{We figured out that these hundred accession numbers are [incorrect] - the quality is not suitable for downstream experiments}, \enquote{I have my doubts that this is actually COVID-19} - these kind of things.}

\textbf{Knowing what biases may exist in your dataset} - One participant wishes for machine-readable bias metadata for datasets: 

\blockquote{[It would be useful if] machines know what part of the dataset you have access to or not. For example, what kind of biases are supposed? If you don't have access to one hundred percent of the data, if you only have access to 70\%?}

\textbf{Other elements of trustworthy data} that participants called out include intentional and clear null values (that can be expressly discerned from accidentally forgotten data gaps), consistent naming conventions for files and entities, and collaboration between the humans who create these datasets.

Finally, a participant highlights that a small dataset which meets multiple quality standards (a dream datasource, rather than a real one) is far more valuable than a large but messy data set:

\blockquote{With a few variables, you can do so much, and with more variables, you can, of course, do a lot more. But to me, my main pain point was this integration, harmonization and worry about quality. [...] The last thing you want is to publish something, and it's not interpreted correctly or it's lacking, or we missed something. That scares me, and so my dream data would be quality in terms of well-done integration, harmonization, standardization, of the data and if that's done, even if small, I would take it, and be happy with that.}

\subsubsection{Datasource quality may change or improve over time, and no data are ever perfect}

\blockquote{I try to steer away from examples that kind of held up as an \enquote{ideal use case}, because those don't tend to exist.}

Most of the good or "dream" attributes of data sources shown above are individual aspects of imperfect datasets - no one dataset combines all these attributes into a single neat package. Over time, however, around one third of participants report that some data sources improved in response to pandemic pressure. Others reported that data sources \textit{did} change, but not for the better. 

One participant describes taking a year to analyse and clean a clinical data set, and create a computational pipeline to easily produce analyses with the data in the future: 

\blockquote{So [after] slightly more than a year, we've gone through the manual [data], and we've also gone through the automated [data]. We have a tool now and we have the data that goes through the computational pipeline. I think there's a big difference in the quality of that data compared to the original. In fact, we also identified issues that were maybe not obvious unless you had done those detailed analyses. So all in all, I think this was a good experience and a good exercise. The teams are very happy that there's actually a tool [...] they're excited to be able to use this tool for their other research interests.}

Another participant notes that digitisation of government services has improved in response to the pandemic: 

\blockquote{The digitalisation of government services has been a long and overdue agenda here in [our region]. We say all governments to some degree, managed to improve at least a bit, the way they engage online because of the pandemic.}

\subsection{Risks and Ethics in COVID-19 data sharing }

\begin{figure}[h!]
    \centering
    \includegraphics[width=1\linewidth]{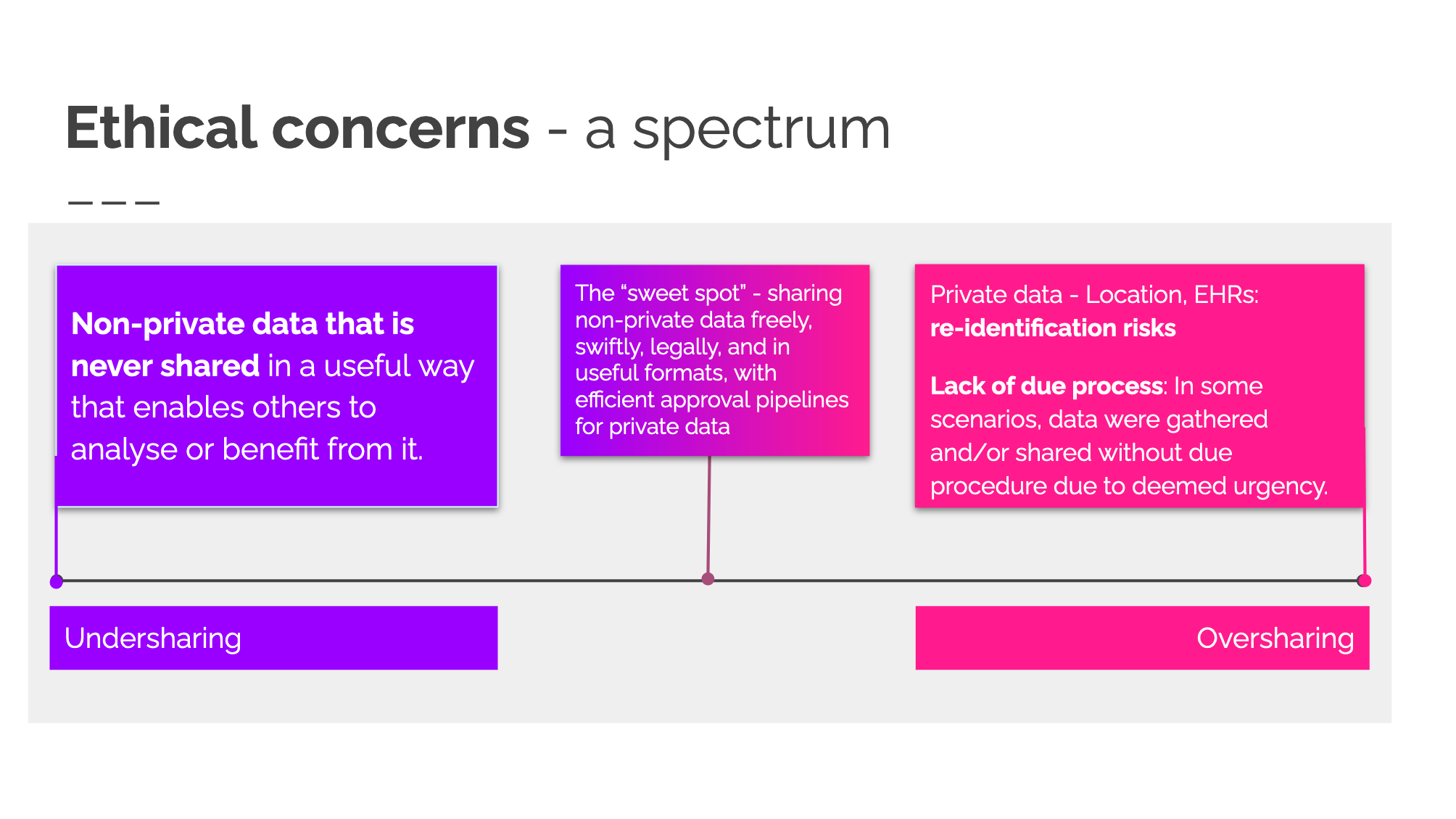}
    \caption{Primary ethical concerns from participants ranged from under-sharing life-saving data, up to over-sharing data that \textit{should} have remained private}
    \label{fig:fig-ethics}
\end{figure}

We asked interview participants whether they had any ethical issues they wished to highlight. We do not report in detail on this section, as in most cases, there was little additional content that had not been highlighted in previous sections of the interview, such as the ethical imperative to share information smoothly in order to reduce the impact of pandemic waves. The need to share was occasionally balanced by the experience of participants who'd been allowed access to patient data so swiftly they doubted that appropriate consent had been sought or granted. We view this range of concerns as a spectrum, shown in Figure \ref{fig:fig-ethics} with a difficult-to-find "sweet spot" in the middle, where data are shared in a manner that allows use without breaching ethical and privacy concerns.

Equity was also raised as a worry; one participant from a high income region noted that they wanted to deal equitably with lower resource settings, but were unsure how to best do this without being patronising, colonialist, or extractive. On the other hand, participants from lower-resource settings did not raise extractivism and colonialism as concerns at all. One participant from a lower-resource setting expressly wished that people from a higher-income setting would treat low-resource setting generated research data as valuable, useful, and something to learn from. We note that these are individual points of view, and unlikely to be highly representative of others in similar settings. 

\section{Discussion}

When the COVID-19 pandemic began to affect countries globally in early 2020, governments, funding bodies, researchers, technologists, medical professionals, and many others responded in en masse to the crisis. In some cases, existing data infrastructure was ready for the demands placed upon it, either because it had been designed with an epidemic in mind, or because its human, social, legal, and technical frameworks were designed to be swiftly adapted. In many other cases, the frameworks to cope with a sudden urgent need for data were insufficient or nonexistent. 

Given that human throughput may be put under sudden and unplanned pressure due to crises such as a pandemic or epidemic, an emphasis should be placed on designing flexible knowledge management systems that can swiftly be adapted. Participants in this study reported in multiple instances that some of the best responses they had seen were in scenarios where there was such an infrastructure - technical (such as a medical reporting system that could easily be adapted to add new data columns) or human (such as "data champions" embedded in organisations). These data-sharing-ready infrastructures and cultures were able to rapidly pivot and focus their tools or skills on preparing better pandemic-relevant data, \textit{even though the systems themselves were not designed with pandemic data in mind}. This is also consistent with \cite{gil-garcia_government_2016}'s observations that dedicated data-sharing project managers were an essential component of effective inter-agency governmental data sharing.

Some of the barriers highlighted in the previous section are unsurprising, and may have relatively straightforward solutions that primarily demand dedicated resourcing to overcome. Indeed, if data standards that "best practices" researchers have been pushing for (such as FAIR data) were implemented more widely, many of the barriers discussed in this paper would not exist. In other cases, however, the combination of social, commercial, ethical, or legal barriers may conflict and require trade-offs, or change may be needed at a more systemic and time-consuming level. 

The need for complete datasets and contextual metadata in private records presents a quandary: in contrast to non-private data, federated or linked queries across multiple private or semi-private data sources presents risk of de-anonymisation that is often deemed too high-risk or complicated, although there are initiatives such as trusted research environments that do attempt this. One participant (section 5.3.2.1) highlighted that end-to-end medical records would provide significant value both to research and to personal medical outcomes, providing context that otherwise would be missing from "the partial observation that we get in medical records".

Getting as complete a picture as possible of data and its context is imperative if data analysts are to create valid interpretations. Existing literature shows that researchers may be reluctant to share their data unless it was accompanied by contextualised metadata, for fear of their data being misinterpreted  (\cite{datasharing_rcts}, \cite{Yimei_Zhu_Open_access_in_uk}, \cite{empirical_datasharing_plos}). Despite this, a recurring theme in many of the data-use barriers (Barrier 3) was that contextual information may be missing from raw data sets, possibly due to lack of time and resources to create the context (Barrier 5). 

\subsection{Geopolitical data suppression, or following the science? To be effective, policy \textit{must} be informed by evidence}

An oft-cited pandemic mantra is "follow the science". This assertion perhaps lacks the nuance to recognise that many scientific pandemic responses will require trade-offs between different pressures and needs, but one thing is clear nevertheless: without transparent and effective sharing of pandemic related data, it becomes highly challenging to follow the science, as science will be missing a significant part of the picture. 

Where data are provided by government bodies or by extensions of government bodies (such as national government-sponsored healthcare and schooling), data sharing and transparency can become political. Multiple participants, from multiple different countries, reported challenges, redactions, and inconsistencies around governmental data sharing in their region, and speculated upon multiple different reasons. It is usually unclear whether some of these gaps were accidental, perhaps due to human throughput challenges (Barrier 5), or whether they were part of a small-scale or concerted government efforts to obscure or ignore the effects of the pandemic upon hospitals, schools, and other infrastructure. More than one participant talked about government-agency-generated or scientific data they had seen with their own eyes which painted a grim picture of COVID-19 transmission or evolution, which was never published, was published in aggregate form that obfuscated serious problems, or which later disappeared without explanation. 

As long as short-term political reputation is prioritised over transparency, governmental and healthcare responses to the pandemic will be insufficient, resulting in preventable death and chronic illness - a terrible humanitarian disaster and long-term economic burden that could otherwise have been avoidable.

\subsection{A more severe crisis for those who have the least: Equity in pandemic data and pandemic response}

The COVID-19 pandemic has also magnified inequity amongst the most disadvantaged (\cite{alon_gender_inequality_2020}, \cite{pilecco_racial_inequality_2020}). Multiple participants highlighted that datasets may not equitably represent populations (biased data), and/or may not have sufficient data to be able to tell if different populations are experiencing equitable outcomes (potentially biased data, with no easy way to tell if the bias is present or not). This once again underscores the importance of this study's recurring theme of needing contextual data and metadata around pandemic data, such as socioeconomic status, gender, race/ethnicity, etc. when designing human-centric systems for epidemic and healthcare data. 

Participants in the study reported that open data - even if lower quality - is more valuable to them than data that had restrictions on access, use, or re-sharing. \cite{chen_ethical_ml_2021} highlighted that opportunistic use of non-representative data often serve to exaggerate and reinforce equity biases in data. Whilst some participants in the study were well aware that such "convenience sampling" may result in biases, they had relatively little control to improve this in any way, as the other choice would be not to perform their analysis at all. 

\subsection{Law as code: Machine-readable legal standards}

One of the most unexpected findings of this study was the importance of highly detailed temporal legal metadata when modelling pathogen spreads. Sections 5.2.3.5 to 5.2.3.7 highlight that data changes over time, and when geographical governmental boundaries change, knowing \textit{when} a geographical file is produced will affect how the data are interpreted. Similarly, modelling pathogen spread based on recent and historical evidence requires knowledge of what interpersonal mingling rules were in place at the time and in the days or weeks beforehand. These rules might include whether a lockdown was in place, whether mask wearing was mandatory, whether small or large groups could legally mingle, and whether work-from-home or school-from-home rules were in force. Spikes and lulls in metrics such as tests, positive cases, hospitalisations, and death rates become hard to explain when the legal behavioural context around the spread isn't known. 

Whilst the simplest solution to this type of quandary is a human-readable date-based list of temporal law and/or geographical boundary changes, there is also scope for an even more powerful implementation that would better aid computational modellers seeking to interpret and predict pathogen spread: versioned and timestamped machine-readable laws. This combines the need for temporal metadata with best practice for data access, would fulfil the participant wishlist for programmatic access to data (section 5.3.2.2), would be better compliant with FAIR data guidelines, and is supported by previous literature, such as the New Zealand Better Rules for Government Discovery \cite{better_rules_machine_legislation}, which asserts: 

\blockquote{The models developed to create [machine consumable legal] rules can be reused: to contribute to modelling
and measurement of the real impact of the policy/rules on people, economy and
government.}

A careful implementation of this technique could be immensely powerful, enabling computational spread modellers and policy makers to collaborate on powerful computational pathogen spread models. These models could effectively explain and understand previous pandemic data whilst also predicting the effects of different pathogen-related interpersonal mingling laws by updating the legal parameters in a test environment and allowing this to feed into existing computational models automatically. Rapid feedback loops between existing data and potential outcomes will save lives, reduce long-term disability, and reduce undesirable economic consequences. 

\section{Conclusion and future work}

\begin{figure}[!ht]
    \centering
    \includegraphics[width=1\linewidth]{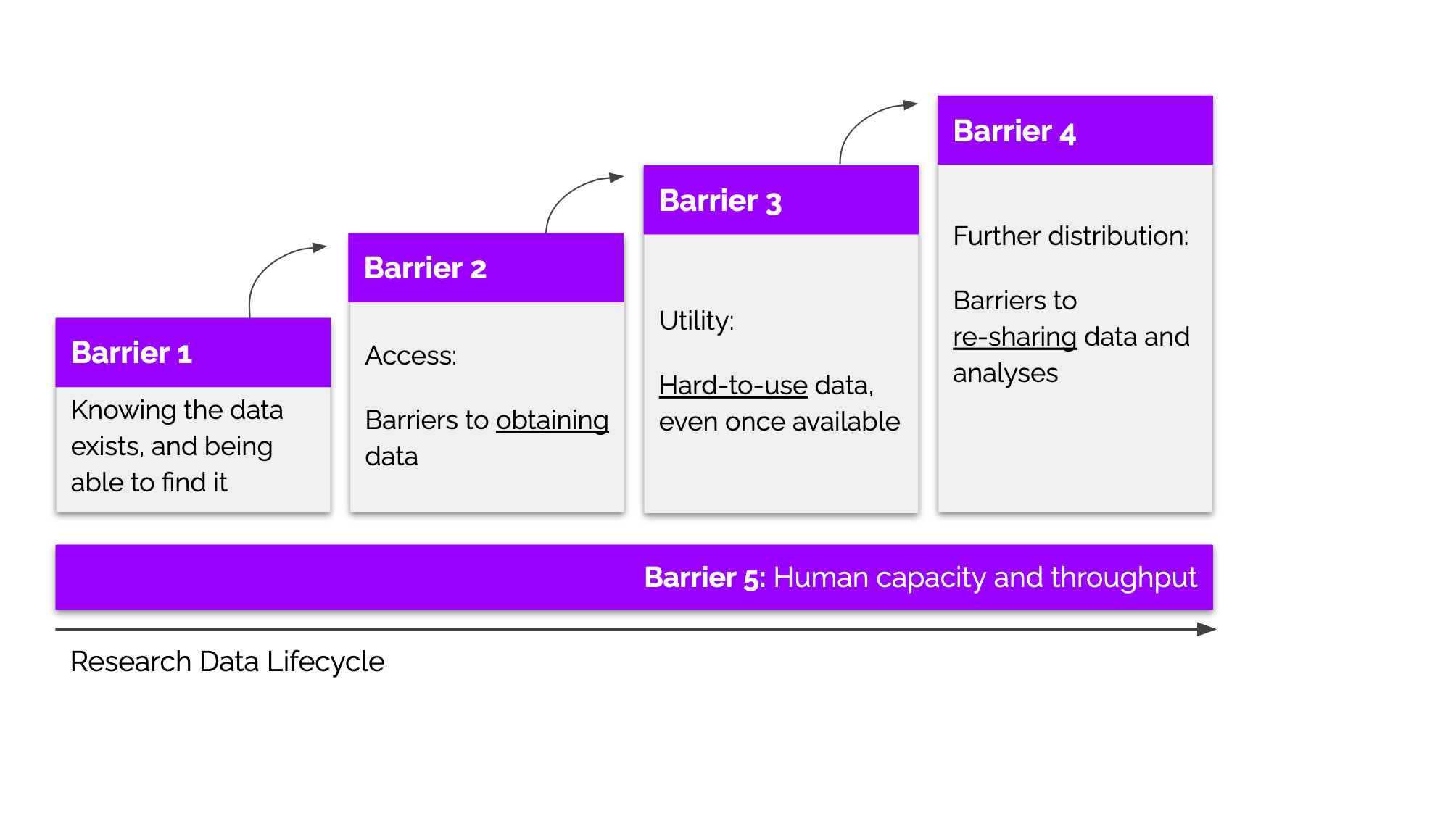}
    \caption{Barriers 1 through 5 are sequential, and all are cumulative. 1: Knowing data exists; 2: Access to data; 3: Utility of the data; 4: Further distribution. Barrier 5 underpins the other four: Human throughput}
    \label{fig:fig-barriers}
\end{figure}

Whilst reducing the barriers discussed here would likely make future pandemic responses more effective, the COVID-19 pandemic highlighted issues that already existed. By and large, it seems unlikely that creating more robust data sharing systems, computer-readable legal codes, 
The barriers we discussed throughout this paper are known and supported by existing data sharing and data standards literature and calls for reform, this paper highlights specifically how they apply to the specific crises of pandemics and epidemics. In particular, the need for machine-readable temporal context for legal movement restrictions and geographical data files represents a new and meaningful contribution to existing knowledge. Similarly, it is important to draw attention to the fact that data are still hidden for the sake of political gain, even in a global and long-term humanitarian crisis. 

\textbf{Figure \ref{fig:fig-barriers} }shows a summary of the barrier types and their sequential and cumulative nature over the life cycle of a research project. 

\subsection{Planning for better responses in the future}
A famous adage about investing in the future is that the best time to plant a tree is twenty years ago; the second-best time is now. Many of the barriers participants reported might have been mitigated or reduced, had measures around data sharing and data access been implemented \textit{before} a public health crisis put significant pressure on our existing healthcare, health surveillance, and research systems. Whilst we can not collectively go back and put better data sharing systems in place, we can use the lessons from this pandemic to prepare better for future waves and for other similar crises. 

Data-finding, access, use, sharing, and re-sharing barriers may occur sequentially rather than concurrently, delaying useful responses to urgent health emergencies. Data access requests and grant applications typically turn around in weeks or months. If a pandemic modeller must first apply for a grant to purchase access to data, and only later request access to the data \textit{if} funding is approved, one or many pandemic waves may pass before researchers can create realistic prediction models and recommended mitigation acts. 

Governments and research institutes worldwide are recognising the need for better data-related infrastructure. Recent examples of this are the FAIR COVID-19 data recommendations for a coordinated response by \cite{FAIR_data_for_a_coordinated_COVID-19_response},  and the Goldacre review (\cite{goldacre_review}), which cites cultural, funding, technical, and privacy-aware recommendations for creating more effective healthcare-related data infrastructure. Implementing report recommendations such as these would dismantle many of the barriers found in this research, and improve the lot of healthcare research and outcomes significantly.

This study was carried out during the depths of lockdowns and early COVID-19 vaccine rollouts. As of late 2023, transmission tracking has largely ceased, and mitigation measures have largely disappeared globally. The vaccine prevents the most severe illness, but immunity wanes swiftly, and the virus mutates swiftly enough that some individuals have had two, three, or more cases of the virus (\cite{collaboratory_burden_2022}). All observed infection cases - even ones where the acute phase is “mild” - result in invisible cardiovascular damage, even in previously healthy people, and repeated infections increase risk of the long-term consequences known as “Long COVID” (\cite{davis_et_al}). COVID-19 is simultaneously a serious public health crisis, likely to increase the levels of long-term disability, yet also being ignored by governments, health policy, and the populace as “over”. 

All this suggests that worldwide, nations do not attempt to “follow the science”, as some may have previously claimed. Further research in the area of systemic barriers and data quality control might be useful. Given that we already have a large amount of scientific evidence that is not being acted upon, however, we would argue that the most effective future work here might be policy work. Research on real-world problems seems to be of little point if the knowledge is not used to create real-world solutions. 

The human cost of COVID-19 so far has been staggering. If we wish to ensure that we are in a position to mitigate future epidemics and pandemics, we must collectively dismantle systemic barriers, and build policies which not only enable efficient re-use of non-private data, but also ensure the data itself is intentionally designed to be accessed computationally with a minimum of human input, reducing costly errors and deadly bottlenecks to informed decision making. 

\begin{Backmatter}
%TC:ignore

\paragraph{Acknowledgements}
This paper has been deposited on arXiv as a preprint, DOI: 10.48550/arXiv.2205.12098

A preliminary version of this research was presented at Data for Policy 2021 as an extended abstract, DOI: 10.5281/zenodo.5234417. 

Yo Yehudi would like to thank Kirstie Whitaker for study design and ethics support, and the open science, open data, and OLS community in particular for helping to spread word and recruit for the study.

\paragraph{Competing Interests}
None

\paragraph{Funding Statement}
This work was supported in part by grant EP/S021779/1.

\paragraph{Data Availability Statement}
Data for this study are stored on University of Manchester infrastructure, but are not available or deposited publicly in order to protect the privacy of the individuals who were interviewed.

\paragraph{Ethical Standards}
The research was approved by the University of Manchester and meets all ethical guidelines, including adherence to the legal requirements of the study country.

\paragraph{Author Contributions}
Y.Y. designed the study, gathered and analysed the data, wrote the first draft, and approved the final version of the manuscript. L.H.N. analysed the data and approved the final version of the manuscript. C.J and C.G. supervised ideation, study design, revised the manuscript and approved the final version. 

\paragraph{Supplementary Material}
All quotes have been included as a supplementary file, listed in the order shown in this text.

\bibliographystyle{unsrtnat}
\bibliography{references}

% \IfFileExists{./covid19DataSharing.bbl}{
%   \input{document.bbl}
%   %
%   % manually balance the columns on the final page by increasing right column, due to phantom bibliography preventing automatic balancing
%   \atColsEnd{\vskip-455pt}
%   % force dependency detection within the submission system to find "references.bib", s.t. usebib can find it
%   \suppress%
%   \begingroup%
%   % do not create another section; from http://tex.stackexchange.com/a/22654
%   \renewcommand{\section}[2]{}%
%   \bibliography{references}
%   \endgroup%
%   \endsuppress%
% }{
%   \bibliography{references}
% }

%TC:endignore
\end{Backmatter}

\end{document}